\documentclass[10pt,conference]{IEEEtran}
\IEEEoverridecommandlockouts
\usepackage{longtable}
\usepackage[noadjust]{cite}
\usepackage{amsmath,amssymb,amsfonts}
\usepackage{algorithmic}
\usepackage{graphicx}
\usepackage{comment}
\usepackage{textcomp}
\usepackage{xcolor}
\usepackage{adjustbox}
\usepackage{textcomp}
\usepackage[numbered]{bookmark}
\usepackage{soul}
\usepackage{booktabs}
\usepackage{multirow}
\usepackage{float}
\usepackage{tcolorbox}
\usepackage{graphicx}
\usepackage{url}
\usepackage{tabularx} 
\usepackage[justification=centering]{caption}
\usepackage[switch]{lineno}
\usepackage{ragged2e}
\usepackage{dirtytalk}
\usepackage{multicol}
\usepackage{csquotes}
\usepackage{pgfplots, pgfplotstable}
\usepackage{pgf-pie}
\usepackage{colortbl}
\usepackage{xspace}
\newcommand{\ie}{\textit{i.e., \xspace}}
\newcommand{\eg}{\textit{e.g., \xspace}}
\newcommand{\etal}{\textit{et al. \xspace}}
\usepackage{array}
\usepackage{balance}
\usepackage{supertabular} 
\usepackage[flushleft]{threeparttable}
\usepackage{supertabular}

\newcolumntype{L}{>{\arraybackslash}m{16cm}}
\newcolumntype{C}[1]{>{\centering\let\newline\\arraybackslash\hspace{0pt}}m{#1}}
\newcolumntype{R}[1]{>{\raggedleft\let\newline\\arraybackslash\hspace{0pt}}m{#1}}
\def\BibTeX{{\rm B\kern-.05em{\sc i\kern-.025em b}\kern-.08em
    T\kern-.1667em\lower.7ex\hbox{E}\kern-.125emX}}
    
    \newcommand{\RQone}{What problems are typically perceived by students as true positives versus false positives?\xspace}
\newcommand{\RQtwo}{What category of problems typically takes longer to be fixed?\xspace}
\newcommand{\RQthree}{What is the perceived usefulness of PMD?\xspace}
\begin{document}

\title{\huge On the use of static analysis to engage students with software quality improvement: An  experience with PMD\\
}

\newcommand{\eman}[1]{\textcolor{violet}{{\it [Eman says: #1]}}}
\newcommand{\ali}[1]{\textcolor{blue}{{\it [Ali says: #1]}}}
\newcommand{\mohamed}[1]{\textcolor{red}{{\it [Mohamed says: #1]}}}

\author{
\IEEEauthorblockN{Eman Abdullah AlOmar\IEEEauthorrefmark{1}, 
Salma Abdullah AlOmar,
Mohamed Wiem Mkaouer\IEEEauthorrefmark{2}}
\IEEEauthorblockA{\IEEEauthorrefmark{1}Software Engineering Department, Stevens Institute of Technology, Hoboken, NJ, USA\\
\IEEEauthorrefmark{2}Software Engineering Department, Rochester Institute of Technology, Rochester, NY, USA\\ 
\text{ealomar@stevens.edu}, \text{alomar.salma@gmail.com}, \text{mwmvse@rit.edu}\\
}}
\maketitle

\begin{abstract}
\textcolor{black}{Static analysis tools are frequently used to scan the source code and detect
deviations from the project coding guidelines.}
 \textcolor{black}{Given their importance, linters are often introduced to classrooms to educate students on how to detect and potentially avoid these code anti-patterns}. However, little is known about their effectiveness in raising students’ awareness, given that these linters tend to generate a large number of false positives. 
  \textcolor{black}{To increase the awareness of potential coding issues that violate coding standards, in this paper, we aim to reflect on our experience with teaching the use of static analysis for the purpose of evaluating its effectiveness in helping students with respect to improving software quality.}
  This paper discusses the results of an experiment in the classroom, over a period of 3 academic semesters, involving 65 submissions that carried out code review activity of 690 rules using PMD. The results of the quantitative and qualitative analysis show that the presence of a set of PMD quality issues influences the acceptance or rejection of the issues, design, and best practices-related categories that take longer time to be resolved, and students acknowledge the potential of using static analysis tools during code review. Through this experiment,  code review can turn into a vital part of the educational computing plan. We envision our findings enabling educators to support students with code review strategies in order to raise students’ awareness about static analysis tools and scaffold their coding skills.

\end{abstract}


\section{Introduction}
\label{Section:Introduction}

Linting is the process of using static analysis tools to scan the source code and detect coding patterns that are considered bad programming practices. These patterns can be responsible for future bugs and stylistic anomalies beyond compiler errors. Given their importance, linters have been introduced in classrooms to educate students on detecting and potentially avoiding these code anti-patterns \cite{mengel1999case}. 
 However, little is known about their effectiveness in raising students' awareness with respect to anti-patterns, given that these linters tend to generate a large number of false positives \cite{johnson2013don,vassallo2020developers,christakis2016developers,sadowski2018lessons}.

In this paper, we reflect on the experience of using linters to support students with their task of debugging and improving the quality of existing systems. In particular, we require students to use PMD \footnote{https://pmd.sourceforge.io}, a state-of-the-art static analysis tool, to detect potential issues in a software system that they did not implement themselves, and then, for each reported issue, they reason whether it should be corrected and suggest corrective action, in the form of a code change, depending on the type of issue reported. The pedagogical goals of this assignment are multiple: 1) Develop the skill of enhancing source code quality through static analysis. Students will be exposed to various bad programming practices that they need to reason on how to address them and suggest corresponding fixes. 2) Train students to review existing code, using the linter, reason over its warnings, and only propose a solution if they are convinced. It trains them to contextualize the problem within the code scope and document the decision of whether it has to be fixed. 3) Initiate students with reading and comprehending code that is not theirs. It prepares them for a more realistic industrial setting, where they will eventually be reading and updating existing code bases.

This paper contributes to the broader adoption of static analysis warnings by (i) designing a practical assignment \textcolor{black}{for improving the quality of software systems}, and (ii) reporting experience of using the PMD tool in a software \textcolor{black}{quality assurance course that has been taken by 65 graduate students}. As part of this paper’s contributions, we provide the assignment description and the tool documentation for educators to adopt and extend \footnote{https://smilevo.github.io/self-affirmed-refactoring/}.

The remainder of this paper is organized as follows: Section \ref{Section:Background} reviews the existing studies related to automated static analysis tools. Section \ref{Section:methodology} outlines our experimental setup in terms of data analysis and research questions. Section \ref{Section:Result} discusses our findings, while the reflection is discussed in Section \ref{Section:Reflection}. Section \ref{Section:Threats} captures any threats to the validity of our work, before concluding with Section \ref{Section:Conclusion}.

\section{Related Work}
\label{Section:Background}

\begin{table*}
  \centering
	 \caption{Related work in automated static analysis tool (ASAT).}
	 \label{Table:Related_Work_in_Static_Analysis_Tool}
\begin{adjustbox}{width=1.0\textwidth,center}
\begin{tabular}{lclllll}\hline
\toprule
\bfseries Study & \bfseries Year  &  \bfseries Context & \bfseries Tool & \bfseries Purpose   \\
\midrule
Kim \& Ernst \cite{kim2007warnings} & 2007 & Bug fix & PMD/FindBugs/JLint 
& Study warning prioritization  \\
Plosch \etal \cite{plosch2008relation} & 2008 & Quality & PMD/FindBugs & Study relation between EQA and ASAT  \\
Di Penta \etal \cite{di2009life} & 2009 & Security & Splint/Rats/Pixy & Observe evolution and decay of vulnerabilities  \\
Panichella \etal \cite{panichella2015would} & 2015 & Code review & CheckStyle/PMD & Study if ASAT helps with code review  \\
Beller \etal \cite{beller2016analyzing} & 2016 & Defect classification & CheckStyle/PMD/FindBugs/JSl & Analyse state of ASAT  \\
& & & Eslint/Jscs/Jshint/Pylint/Rubocop \\
Singh \etal \cite{singh2017evaluating} & 2017 & Code review & PMD & Study ASAT helps reducing review efforts  \\
Liu \etal \cite{liu2018mining} & 2018 & Bug fix  & FindBugs & Mine fix patterns for FindBugs violations \\
Digkas \etal \cite{digkas2018developers} & 2018 & Technical debt & SonarQube & Fix issues \& pay back technical debt  \\
Querel \& Rigby \cite{querel2018warningsguru} & 2018 & Bug prediction & FindBugs/JLint & Integrate statistical bug models with ASAT  \\
Marcilio \etal \cite{marcilio2019static} & 2019 & Bug fix & SonarQube & Study how developers use SonarQube  \\
Aloraini \etal \cite{aloraini2019empirical} & 2019 & Security & Rats/Flawfinder/Cppcheck & Study warnings generated by ASAT  \\
& & & PVS-Studio/Parasoft/Clang \\
Trautsch \etal \cite{trautsch2020longitudinal} & 2020 & Code review & PMD & Study the effect of PMD on quality  \\
 Romano \etal \cite{romano2022static} & 2022 & Test-driven development & SonarLint & Study if ASAT affects software quality  \\
Licorish \& Wagner \cite{licorish2022combining} & 2022 & Bug fix & PMD & Detect performance faults \\
 \textbf{This work} & & \textbf{Education} & \textbf{PMD} & \textcolor{black}{Support students in enhancing quality with PMD} \\
 
\bottomrule
\end{tabular}
\end{adjustbox}
\vspace{-.3cm}
\end{table*}

Research on automated static analysis tools (ASAT) has been important to practitioners,  researchers, and educators. The research community has spent considerable effort studying static analysis tools from different domains. This literature has included the usage of ASATs in the context of bug fixes \cite{kim2007warnings,liu2018mining,marcilio2019static}, quality \cite{plosch2008relation}, security \cite{di2009life,aloraini2019empirical}, code review \cite{panichella2015would,singh2017evaluating,trautsch2020longitudinal}, defect classification and predication \cite{beller2016analyzing,querel2018warningsguru}, and technical debt \cite{digkas2018developers}. However, except for \cite{mengel1999case,edwards2019can,luukkainen2022aspa,senger2022helping}, most of the above studies focus on studying and improving the effectiveness of using ASAT for open-source communities, as opposed to our work that focuses on educating students on locating and fixing software defects. In this section, we are only interested in research related to using ASAT. We summarize these approaches in Table~\ref{Table:Related_Work_in_Static_Analysis_Tool}.

Kim and Ernst \cite{kim2007warnings} investigated the possibility of leveraging the removal times for ASAT warning prioritization by utilizing commit histories of ASAT warnings. Later, Plosch \etal \cite{plosch2008relation} correlated software quality metrics and defects with warnings found by various ASATs. The authors utilized three releases of the eclipse ecosystem and demonstrated correlations for various aspects, including size, complexity, and object-oriented software metrics.

In a security-related context, Di Penta \etal \cite{di2009life} performed an empirical study to extract the history of three open-source projects and analyze security-related ASAT warnings using three static code analyzers. In a similar context, Aloraini \etal \cite{aloraini2019empirical} analyzed security-related ASAT warnings using 116 open-source projects. Both of these studies concluded that the warning density of security-related ASAT remains constant throughout their selected time span.

On the other hand, Beller \etal \cite{beller2016analyzing} empirically investigated the usage of ASAT in open-source projects by focusing on the prevalence of ASAT and the evolution of the configurations for different programming languages. In another study, Querel and Rigby \cite{querel2018warningsguru} utilized FindBugs and Jlint for bug prediction. Their main finding revealed the information provided by the ASAT warnings could improve statistical bug prediction models. Liu \etal \cite{liu2018mining} explored ASAT warning over time by performing a large-scale study using the tool FindBugs via SonarQube. Their approach identified fix patterns that are then applied to unfixed warnings. In another study, Digkas \etal \cite{digkas2018developers} utilized SonarQube to detect ASAT warnings and their removal strategies. The authors focused on technical debt and the resolution time assigned by SonarQube to each detected ASAT warning. Marcilio \etal \cite{marcilio2019static} concentrated on developer usage of ASAT through SonarQube. They focused on the active engagement of developers when fixing different types of issues reported by ASAT.  In a similar context, Licorish \& Wagner \cite{licorish2022combining} combined GIN and PMD for code improvements by focusing on detecting performance faults from Stack Overflow code snippets. Their findings show that  static analysis techniques could be combined with program improvement methods to enhance publicly available code.

Some ASATs are used in the context of code review. Panichella \etal \cite{panichella2015would} studies if ASAT warnings are taken care of during the code review process. Their main finding indicated that the density of warnings slightly varies after each code review. Singh \etal \cite{singh2017evaluating} evaluated how ASAT can reduce code review effort. They investigated the overlap between reviewer comments on GitHub pull requests and warnings from the tool PMD. Their finding showed that PMD overlapped with around 16\% of reviewer comments. 
Trautsch \etal \cite{trautsch2020longitudinal} performed a longitudinal study of ASAT warning evolution and the effect of ASAT on quality. The authors analyzed the commit history of 54 projects, taking into account 193 PMD rules and 61 PMD releases. They found that significant global changes in ASAT warnings are mostly related to coding style changes. Another study relevant to our work is by  Romano \etal \cite{romano2022static}.  The authors \cite{romano2022static} studied the benefits of leveraging an ASAT on software quality in the context of test-driven development (TDD). Their study reveal that the use of a SonarLint helps the participants to improve software quality, although the participants found that TDD is more difficult to be performed. 

To summarize, the study of static analysis tools has been extensively studied (\eg \cite{romano2022static}\cite{liu2018mining} \cite{plosch2008relation}). Since we are focusing on Java,  there are a few widely adopted Java-based open-source static analysis tools such as CheckStyle \footnote{https://checkstyle.sourceforge.io}, FindBugs \footnote{http://findbugs.sourceforge.net}, JLint \footnote{http://jlint.sourceforge.net/}, PMD \footnote{https://pmd.sourceforge.io} and SonarQube \footnote{https://www.sonarqube.org/}. The choice of PMD is motivated by different factors:  its widespread use in the Java community and its maturity (\ie it is available since 2002 and therefore has been in use for a long time \cite{trautsch2020longitudinal}), and works on the Java source code to find coding style problems, which is not the case by FindBugs and JLint that work on byte code and focus on finding programming errors and neglecting programming style issues \cite{plosch2008relation}. CheckStyle focuses more on readability problems compared to PMD, which tends to highlight suspicious situations in the source code \cite{panichella2015would}. Further, although there are recent studies that explored the use of static analysis tools on open-source communities focusing on how warnings evolve across the software evolution history \cite{trautsch2020longitudinal,di2009life,kim2007warnings}, \textcolor{black}{this is the first study investigating how PMD can support students in improving code quality, and reflecting on how students are using it in the classroom.} To advance the understanding of the practice of learning how to find and fix bugs, in this paper, we performed a study in an educational setting using open-source projects having a large number of issues. This study complements the existing efforts that are done in open source systems \cite{kim2007warnings,plosch2008relation,di2009life,panichella2015would,beller2016analyzing,singh2017evaluating,liu2018mining,digkas2018developers,querel2018warningsguru,marcilio2019static,aloraini2019empirical,trautsch2020longitudinal,romano2022static,licorish2022combining} and in education \cite{mengel1999case,edwards2019can,luukkainen2022aspa,senger2022helping}, by complementing the quantitative analysis with a qualitative one, providing evidence of several kinds of warnings students pay more attention to, during code analysis.




\section{Study Design}
\label{Section:methodology}
\subsection{Goal \& Research Questions}
We formulate the main goal of our study based on the \textit{Goal} \textit{Question} \textit{Metric} template \cite{van2002goal}, as follows:



\begin{tcolorbox}
\textit{\textbf{Analyze} the use of an automated static analysis tool (ASAT) \textbf{for the purpose of} \textcolor{black}{familiarizing students with improving source code, by developing the culture of reviewing unknown code and patching it} \textit{with respect to} software quality \textbf{from the point view} of educators \textbf{in the context of} Master's students in SE/CS who analyze Java-based software projects.}
\end{tcolorbox}

According to our goal, we aim to answer the following research questions:

\begin{itemize}
\item \textbf{RQ1. \textit{\RQone}}

\underline{Motivation:} This RQ aims at evaluating whether students can apply analytical skills to identify and fix issues in existing systems. The findings will shed some light on the feasibility of this learning activity to educate students to perform effective code reviews and quality control.

\underline{Measurement:} We examine (1) the types of issues that can be identified and  (2) the PMD ruleset categories that are perceived by students as a true positive and false positive. 

\item \textbf{RQ2. \textit{\RQtwo}}

\underline{Motivation:} This RQ investigates which PMD ruleset is taking longer time to be fixed, with respect to other rulesets. The finding raises educators' awareness of issues types that are hard for students to understand and address. 

\underline{Measurement:} We examine the resolution time taken by students to fix each ruleset, clustered by category.

\item \textbf{RQ3. \textit{\RQthree}}

\underline{Motivation:} This RQ explores the tool's feedback, and how students perceive PMD in general. The finding will inform educators on the usefulness, usability, and functionality of the tool, and allow them to make decisions about what ASATs can better support students with code improvement.

\underline{Measurement:} We examine students' feedback. We extract all their positive and negative comments as they describe their experience with using PMD.

\end{itemize}
 
 As part of this paper’s contributions, we provide the assignment description, dataset, and tool documentation for educators to adopt and extend\footnote{https://smilevo.github.io/self-affirmed-refactoring/}. 
 
\begin{figure}[t]
 	\centering
 	\includegraphics[width=1.0\columnwidth]{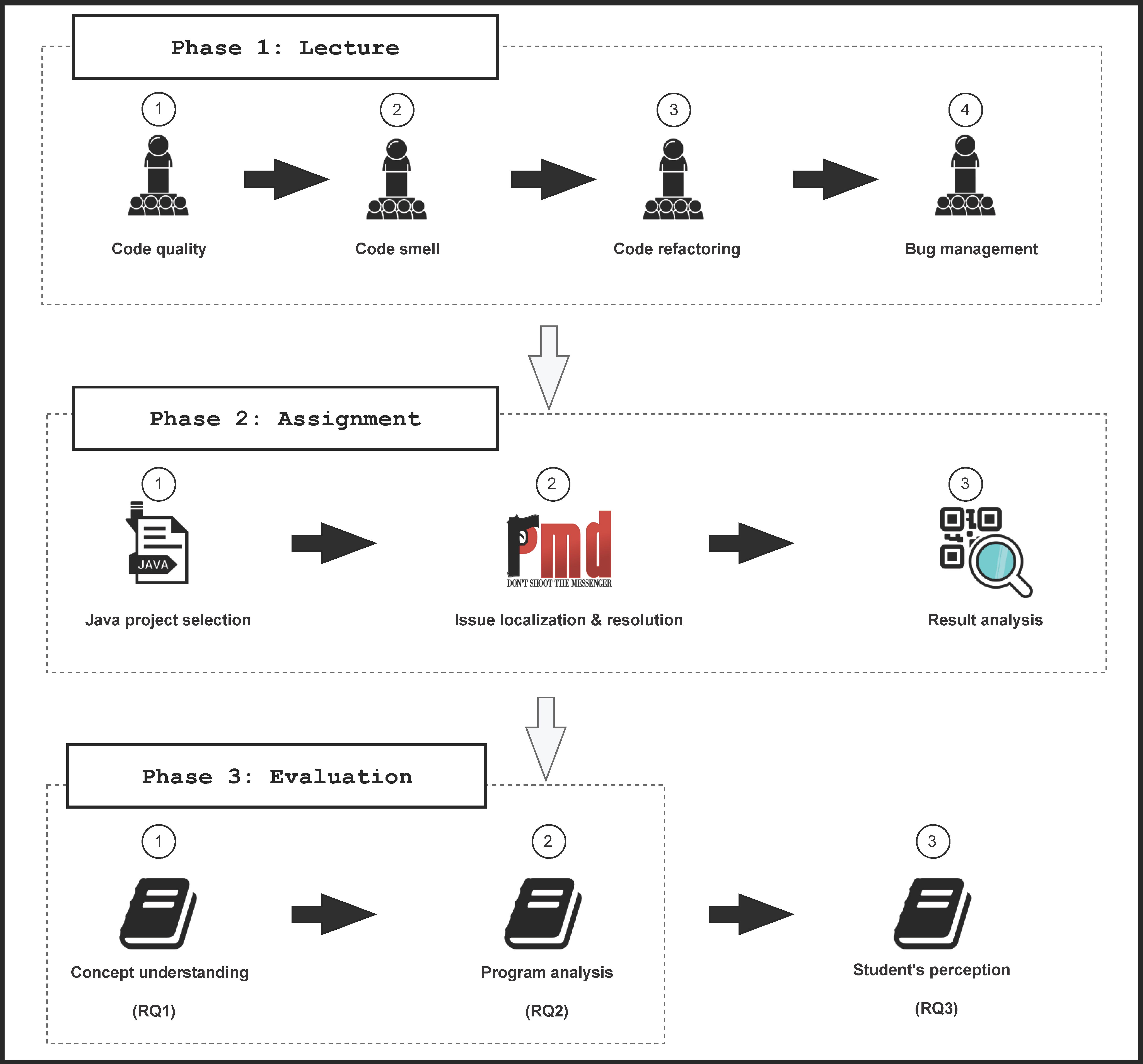}
 	\caption{Key phases of our study.}
 	\label{fig:approach}
\end{figure}

\subsection{Course Overview}

Software quality assurance is a graduate course, consisting of 2 lectures weekly, an hour and 15 minutes each. The course explores the foundations of software quality and software maintenance and introduces several challenges linked to various aspects of software evolution along with support tools to approach them. Also, the course covers various concepts related to software analysis and testing, along with practical tools, widely used as industry standards. Students were also given a number of hands-on assignments related to software quality metrics, code refactoring, bug reporting, unit and mutation testing, and technical debt management. The course deliverables consisted of five individual homework assignments, a research paper reading and presentation, and a long-term group project.

\subsection{Teaching Context and Participants}

The study involves one assignment in the software quality assurance course. The course was taught at Stevens Institute of Technology and Rochester Institute of Technology for a period of 3 semesters. Before conducting the assignment, students have already learned about several code, and design quality concerns: (1) code quality (teaching quality concepts and how to measure software quality), (2) code smells (teaching bad programming practices that violate design principles), (3) code refactoring (teaching refactoring recipes that help improving software quality), and (4) bug management (teaching software bugs and how to locate and fix them). The assignment constituted 7.5\% of the final grade. It was due 14 days after the four corresponding sessions.

\subsection{Assignment Content and Format}

Initially, students are asked to analyze one version of a JAVA software of their choice approved by the instructor to ensure its eligibility based on popularity, besides making sure it correctly compiles, since PMD requires it. The rationale behind giving students the choice of project is to let them choose one that they are comfortable with and that fits into their interests. 
For students who do not want to search for a project, they have given a list that the instructor has already curated (see Table \ref{Table:Project}). 
We selected these projects as we already know they contain a variety of software defects. 
Then, students are requested to set up and run PMD to analyze the chosen project production code. Students are also given the choice of running either the stand-alone version of the tool, or one of its plugins associated with popular IDEs (Eclipse, IntelliJ), as we want the students to be familiar with the coding environment, and reduce the setup overhead. Upon running PMD, students are required to choose a minimum of 10 warnings, and at least one from each category, if applicable. We enforce the diversification of warnings, to ensure a wider exposure to different types of potential issues, varying from design to multithreading, and documentation. It also increases students' learning curve as they cannot reuse their fix to address multiple instances of the same exact warning. Yet, we allow students to choose the instances they want to address. It implicitly makes students read many warnings, from all categories, which increases the likelihood of incidental learning. Furthermore, letting students choose the code fragments to review, increases their confidence in the decision they will make with respect to the given warning, \ie either consider it as a true positive, and provide a code fix, or consider it as a false positive, and provide a justification.
In a nutshell, students followed these steps:

\begin{enumerate}
    \item Install the PMD.
    \item Run PMD on a project of students' choice and select 10 issues of different types.
    \item Report the findings for each issue: (1) the type of issue, (2) whether it is a true or a false positive, (3) if it is a true positive, what are the necessary steps to fix it, (4) how long it took to check it / fix it, and (5) the code snippet.
    \item Add to the report a concise comment about the experience with PMD (optional).
    
\end{enumerate}

Submissions artifacts' evaluation was based on 1) assessment of students’ ability to understand the type of issues (concept understanding (RQ1)); 2) assessment of whether students have provided acceptable fixes, or proper justification in both cases or accepting or rejecting PMD's recommendation (program analysis and evolution (RQ2)). Students' perception of the code was excluded from the evaluation process, as it can bias the experiment, as students would be filling out the survey arbitrarily, under the pressure of being graded. Also, providing feedback was anonymous and not mandatory, to increase the magnitude of PMD usage experience (RQ3). Although feedback was optional, many students have completed it (92.3\%). Figure \ref{fig:approach}  depicts an overview of our experiment setup and execution.


\subsection{Assignment Execution}

The assignment was performed over three consecutive semesters. 65 students, primarily from computer science (CS) and software engineering (SE) majors, were enrolled during these semesters and completed the assignment. 


\subsection{Data Analysis}
\label{subsec:data-analysis}

    We analyzed the responses to the open-ended question to create a comprehensive high-level list of themes by adopting a thematic analysis approach based on guidelines provided by Cruzes~\etal~\cite{cruzes2011recommended}. Thematic analysis is one of the most used methods in Software Engineering literature~\cite{Silva2016why,alomar2022code,alomar2021refactoring}. This is a technique for identifying and recording patterns (or \say{themes}) within a collection of descriptive labels, which we call \say{codes}. For each response, we proceeded with the analysis using the following steps: \textit{i}) Initial reading of the survey responses; \textit{ii}) Generating initial codes (\ie labels) for each response; \textit{iii}) Translating codes into themes, sub-themes, and higher-order themes; \textit{iv}) Reviewing the themes to find opportunities for merging; \textit{v}) Defining and naming the final themes, and creating a model of higher-order themes and their underlying evidence. The above-mentioned steps were performed independently by two authors. One author performed the labeling of students' comments independently from the other author who was responsible for reviewing the currently drafted themes. \textcolor{black}{By the end of each iteration, the authors met and refined the themes to reach a consensus.}
    \textcolor{black}{It is important to note that the approach is not a single-step process. As the codes were analyzed, some of the first cycle codes were
subsumed by other codes, relabeled, or dropped altogether. As the two authors progressed in the translation to themes, there was some rearrangement, refinement, and reclassification of data into
different or new codes. We used the
thematic analysis technique to address RQ3.}
\begin{table}
  \centering
	 \caption{The list of open-source projects used in the assignment.}
	 \label{Table:Project}
\begin{adjustbox}{width=0.48\textwidth,center}
\begin{tabular}{lclllll}\hline
\toprule
\bfseries Project & \bfseries \# commits &  \bfseries  \# contributors & \bfseries Domain   \\
\midrule
Ant & 14,887 & 64 & Java builder \\
GanttProject & 4,361 & 38 & Project management\\
Hutool & 4,074 & 191 & Code design \\
JCommander & 1,009 & 64 & Command line parsing\\ 
JFreeChart & 4,218 & 24 & Data visualization \\
JHotDraw & 804 & 3 & Data visualization \\
Log4J & 12,211 & 137 & Logging\\
Nutch & 3,293 & 46 & Web crawler\\
Rihno & 4,119 & 80 & Script builder \\
RxJava & 6,004 & 289 & Java VM \\
Xerces &  6,463 & 5 & XML parser \\

\bottomrule
\end{tabular}
\end{adjustbox}
\vspace{-.3cm}
\end{table}

\section{Results}
\label{Section:Result}

\begin{table*}
\caption{PMD rules that were perceived by students as true positive and false positive, broken down by category, as well as the percentages of responses for which students accepted/rejected the issues. Cells with 100\% acceptance are highlighted in blue, while cells with 100\% rejection are highlighted in red.}
\label{Table:pmd-tp-fp-both-a}
\resizebox{\textwidth}{!}{
 \begin{threeparttable}
\begin{tabular}{l|llc|l|lllccccc}
\toprule
\multicolumn{1}{l}{\textbf{Category}}& \multicolumn{1}{l}{\textbf{Rule}} & \multicolumn{1}{l}{\textbf{Metric}} & \multicolumn{1}{l|}{\textbf{Ratio}} & \multicolumn{1}{l}{\textbf{Category}}& \multicolumn{1}{l}{\textbf{Rule}} & \multicolumn{1}{l}{\textbf{Metric}} & \multicolumn{1}{l}{\textbf{Ratio}} \\ \hline

\parbox[t]{2mm}{\multirow{60}{*}{\rotatebox[origin=c]{90}{\textbf{Best Practices}}}} 
 
& \multirow{2}{*} {\texttt{AbstractClassWithoutAbstractMethod}}  & True Positive &  \cellcolor[HTML]{0350F8} 100\% & \parbox[t]{2mm}{\multirow{60}{*}{\rotatebox[origin=c]{90}{\textbf{Design}}}} & \multirow{2}{*} {\texttt{AvoidCatchingGenericException}}  & True Positive & \cellcolor[HTML]{0350F8} 100\%  \\ 
& &  False Positive  & 0\%  & & & False Positive  & 0\%  \\ \cline{2-4}\cline{6-8}
 
  & \multirow{2}{*} {\texttt{ArrayIsStoredDirectly}}   & True Positive & 80\% & & \multirow{2}{*} {\texttt{AvoidDeeplyNestedIfStmts}}  & True Positive & \cellcolor[HTML]{0350F8} 100\%  \\ 
& & False Positive  & 20\% & & & False Positive  & 0\% \\ \cline{2-4}\cline{6-8}

& \multirow{2}{*} {\texttt{AvoidGlobalModifier}}  & True Positive &  \cellcolor[HTML]{0350F8} 100\% & & \multirow{2}{*} {\texttt{AvoidRethrowingException}}  & True Positive & \cellcolor[HTML]{0350F8} 100\% \\ 
& & False Positive  &  0\% & & & False Positive  & 0\%  \\ \cline{2-4}\cline{6-8}
 
  & \multirow{2}{*} {\texttt{AvoidMessageDigestField}}   & True Positive & \cellcolor[HTML]{0350F8} 100\% & & \multirow{2}{*} {\texttt{AvoidThrowingNullPointerException}}  & True Positive & \cellcolor[HTML]{0350F8} 100\% \\ 
& & False Positive  &   0\% & & & False Positive  & 0\%   \\ \cline{2-4}\cline{6-8}

& \multirow{2}{*} {\texttt{AvoidPrintStackTrace}}  & True Positive &80\% & & \multirow{2}{*} {\texttt{AvoidThrowingRawExceptionTypes}}  & True Positive & 83.33\%   \\ 
& & False Positive  & 20\% & & & False Positive  & 16.66\%   \\ \cline{2-4}\cline{6-8}
 
  & \multirow{2}{*} {\texttt{AvoidReassigningParameters}}  & True Positive & 46.15\% & & \multirow{2}{*} {\texttt{ClassWithOnlyPrivateConstructorsShouldBeFinal}}  & True Positive & 50\%  \\ 
& & False Positive  & 53.84\%  & & & False Positive  & 50\%    \\ \cline{2-4}\cline{6-8}

& \multirow{2}{*} {\texttt{AvoidUsingHardCodedIP}}  & True Positive &  \cellcolor[HTML]{0350F8} 100\% & & \multirow{2}{*} {\texttt{CognitiveComplexity}}  & True Positive & \cellcolor[HTML]{0350F8} 100\% \\ 
& & False Positive  &  0\% & & & False Positive  & 0\%   \\ \cline{2-4}\cline{6-8}
 
  & \multirow{2}{*} {\texttt{ForLoopCanBeForEach}}   & True Positive & \cellcolor[HTML]{0350F8} 100\% & & \multirow{2}{*} {\texttt{CollapsibleIfStatements}}  & True Positive & \cellcolor[HTML]{0350F8} 100\% \\ 
& & False Positive  & 0\%  & & & False Positive  & 0\%   \\ \cline{2-4}\cline{6-8}

& \multirow{2}{*} {\texttt{JUnit4TestShouldUseBeforeAnnotation}}  & True Positive & 0\% & & \multirow{2}{*} {\texttt{CyclomaticComplexity}}  & True Positive & \cellcolor[HTML]{0350F8} 100\%  \\ 
& & False Positive  & \cellcolor[HTML]{FE0000} 100\% & & & False Positive  & 0\%  \\ \cline{2-4}\cline{6-8}
 
  & \multirow{2}{*} {\texttt{JUnitAssertionsShouldIncludeMessage}}   & True Positive & \cellcolor[HTML]{0350F8} 100\% & & \multirow{2}{*} {\texttt{DataClass}}  & True Positive & \cellcolor[HTML]{0350F8} 100\% \\ 
& & False Positive  & 0\%  & & & False Positive  & 0\%   \\ \cline{2-4}\cline{6-8}

& \multirow{2}{*} {\texttt{JUnitTestContainsTooManyAsserts}}  & True Positive & \cellcolor[HTML]{0350F8} 100\% & & \multirow{2}{*} {\texttt{DoNotExtendJavaLangError}}  & True Positive & \cellcolor[HTML]{0350F8} 100\%  \\ 
& & False Positive  &  0\% & & & False Positive  & 0\%   \\ \cline{2-4}\cline{6-8}

& \multirow{2}{*} {\texttt{JUnitTestsShouldIncludeAssert}}  & True Positive &  0\% & & \multirow{2}{*} {\texttt{ExcessiveImports}}  & True Positive & 33.33\% \\ 
& & False Positive  &  \cellcolor[HTML]{FE0000} 100\% & & & False Positive  & 66.66\%  \\ \cline{2-4}\cline{6-8}
 
  & \multirow{2}{*} {\texttt{LiteralsFirstInComparisons}}  & True Positive & \cellcolor[HTML]{0350F8} 100\% & & \multirow{2}{*} {\texttt{ExcessiveMethodLength}}  & True Positive & \cellcolor[HTML]{0350F8} 100\%  \\ 
& & False Positive  & 0\% & & & False Positive  & 0\%    \\\cline{2-4}\cline{6-8}

& \multirow{2}{*} {\texttt{LooseCoupling}}  & True Positive & \cellcolor[HTML]{0350F8} 100\% & & \multirow{2}{*} {\texttt{ExcessivePublicCount}}  & True Positive & \cellcolor[HTML]{0350F8} 100\%  \\ 
& & False Positive  &  0\%  & & & False Positive  & 0\%  \\\cline{2-4}\cline{6-8}
 
  & \multirow{2}{*} {\texttt{MethodReturnsInternalArray}}   & True Positive & 50\% & & \multirow{2}{*} {\texttt{FinalFieldCouldBeStatic}}  & True Positive & 80\%  \\ 
& & False Positive  & 50\% & & & False Positive  & 20\%    \\ \cline{2-4}\cline{6-8}

& \multirow{2}{*} {\texttt{MissingOverride}}  & True Positive &  0\% & & \multirow{2}{*} {\texttt{GodClass}}  & True Positive & \cellcolor[HTML]{0350F8} 100\%  \\ 
& & False Positive  &  \cellcolor[HTML]{FE0000} 100\% & & & False Positive  & 0\%  \\ \cline{2-4}\cline{6-8}
 
  & \multirow{2}{*} {\texttt{OneDeclarationPerLine}}  & True Positive & 60\% & & \multirow{2}{*} {\texttt{ImmutableField}}  & True Positive & 83.33\%  \\ 
& & False Positive  & 40\% & & & False Positive  & 16.66\%    \\ \cline{2-4}\cline{6-8}

& \multirow{2}{*} {\texttt{PreserveStackTrace}}  & True Positive & 80\% & & \multirow{2}{*} {\texttt{LawOfDemeter}}  & True Positive & 83.33\%    \\ 
& & False Positive  & 20\% & & & False Positive  & 16.66\%   \\ \cline{2-4}\cline{6-8}
 
  & \multirow{2}{*} {\texttt{ReplaceEnumerationWithIterator}}   & True Positive & \cellcolor[HTML]{0350F8} 100\% & & \multirow{2}{*} {\texttt{NPathComplexity}}  & True Positive & \cellcolor[HTML]{0350F8} 100\% \\ 
& & False Positive  & 0\% & & & False Positive  & 0\%  \\ \cline{2-4}\cline{6-8}

& \multirow{2}{*} {\texttt{ReplaceHashtableWithMap}}  & True Positive &  \cellcolor[HTML]{0350F8} 100\% & & \multirow{2}{*} {\texttt{SignatureDeclareThrowsException}}  & True Positive & \cellcolor[HTML]{0350F8} 100\%  \\ 
& & False Positive  &  0\% & & & False Positive  & 0\% \\ \cline{2-4}\cline{6-8}
 
  & \multirow{2}{*} {\texttt{ReplaceVectorWithList}}   & True Positive & 50\% & & \multirow{2}{*} {\texttt{SimplifyBooleanExpressions}}  & True Positive & \cellcolor[HTML]{0350F8} 100\%  \\ 
& & False Positive  & 50\%  & & & False Positive  & 0\%  \\ \cline{2-4}\cline{6-8}

& \multirow{2}{*} {\texttt{SwitchStmtsShouldHaveDefault}}  & True Positive & 75\% & & \multirow{2}{*} {\texttt{SimplifyBooleanReturns}}  & True Positive &\cellcolor[HTML]{0350F8} 100\%   \\ 
& & False Positive  & 25\%  & & & False Positive  & 0\%  \\ \cline{2-4}\cline{6-8}
 
  & \multirow{2}{*} {\texttt{SystemPrintln}}   & True Positive &50\% & & \multirow{2}{*} {\texttt{SingularField}}  & True Positive & \cellcolor[HTML]{0350F8} 100\% \\ 
& & False Positive  &50\%  & & & False Positive  & 0\%  \\ \cline{2-4}\cline{6-8}

& \multirow{2}{*} {\texttt{UnusedAssignment}}  & True Positive & 71.42\%  & & \multirow{2}{*} {\texttt{TooManyFields}}  & True Positive & 66.66\% \\ 
& & False Positive  & 28.57\%  & & & False Positive  & 33.33\%  \\ \cline{2-4}\cline{6-8}
 
  & \multirow{2}{*} {\texttt{UnusedFormalParameter}}   & True Positive & \cellcolor[HTML]{0350F8} 100\% & & \multirow{2}{*} {\texttt{TooManyMethods}}  & True Positive & 75\%  \\ 
& & False Positive  & 0\%  & & & False Positive  & 25\%  \\ \cline{2-4}\cline{6-8}

& \multirow{2}{*} {\texttt{UnusedImports}}  & True Positive & 58.33\% & & \multirow{2}{*} {\texttt{UselessOverridingMethod}}  & True Positive & 80\%   \\ 
& & False Positive  & 41.66\%  & & & False Positive  & 20\%   \\ \cline{2-4}\cline{6-8}
 
  & \multirow{2}{*} {\texttt{UnusedLocalVariable}}   & True Positive & 90\% & & \multirow{2}{*} {\texttt{UseObjectForClearerAPI}}  & True Positive & \cellcolor[HTML]{0350F8} 100\%  \\ 
& & False Positive  & 10\% & & & False Positive  & 0\%    \\ \cline{2-4}\cline{6-8}

& \multirow{2}{*} {\texttt{UnusedPrivateField}}  & True Positive &  \cellcolor[HTML]{0350F8} 100\% & & \multirow{2}{*} {\texttt{UseUtilityClass}}  & True Positive & \cellcolor[HTML]{0350F8} 100\%  \\ 
& & False Positive  &  0\% & & & False Positive  & 0\%  \\ \cline{2-4}\cline{6-8}
 
  & \multirow{2}{*} {\texttt{UnusedPrivateMethod}}   & True Positive & 80\% & & \multirow{2}{*} {\texttt{}}  &  &  
 \\ 
& & False Positive  & 20\%   & & &  &   \\ \cline{2-4}

& \multirow{2}{*} {\texttt{UseAssertNullInsteadOfAssertTrue}}  & True Positive &  \cellcolor[HTML]{0350F8} 100\% & & \multirow{2}{*} {\texttt{}}  &  &  \\ 
& & False Positive  &  0\% & & &   &  \\ \cline{2-4}
 
  & \multirow{2}{*} {\texttt{UseTryWithResources}}   & True Positive & \cellcolor[HTML]{0350F8} 100\% & & \multirow{2}{*} {\texttt{}}  &  &  \\ 
& & False Positive  & 0\% & & &   &   \\ 

\hline  
\hline
\parbox[t]{2mm}{\multirow{45}{*}{\rotatebox[origin=c]{90}{\begin{tabular}[c]{@{}l@{}}\textbf{Code Style}\end{tabular}}}}
 & \multirow{2}{*} {\texttt{AbstractNaming}} & True Positive & \cellcolor[HTML]{0350F8} 100\% &\parbox[t]{2mm}{\multirow{45}{*}{\rotatebox[origin=c]{90}{\textbf{Error Prone}}}} & \multirow{2}{*} {\texttt{AssignmentInOperand}}  & True Positive & 50\%   \\
& & False Positive  & 0\%   & & & False Positive  & 50\%   \\ \cline{2-4}\cline{6-8}

 & \multirow{2}{*} {\texttt{AtLeastOneConstructor}} & True Positive & 70\%& & \multirow{2}{*} {\texttt{AvoidFieldNameMatchingMethodName}}  & True Positive & 66.66\%  \\
& & False Positive  & 30\%  & & & False Positive  & 33.33\%    \\ \cline{2-4}\cline{6-8}

 & \multirow{2}{*} {\texttt{AvoidFinalLocalVariable}} & True Positive & \cellcolor[HTML]{0350F8} 100\%  & & \multirow{2}{*} {\texttt{AvoidLiteralsInIfCondition}}  & True Positive & 80\% \\
& & False Positive  & 0\%  & & & False Positive  & 20\%     \\ \cline{2-4}\cline{6-8}

 & \multirow{2}{*} {\texttt{AvoidPrefixingMethodParameters}} & True Positive & 0\%  & & \multirow{2}{*} {\texttt{CompareObjectsWithEquals}}  & True Positive & 90.90\%  \\
& & False Positive  & \cellcolor[HTML]{FE0000} 100\%  & & & False Positive  &9.09\%     \\ \cline{2-4}\cline{6-8}

 & \multirow{2}{*} {\texttt{BooleanGetMethodName}} & True Positive & \cellcolor[HTML]{0350F8} 100\% & & \multirow{2}{*} {\texttt{ConstructorCallsOverridableMethod}}  & True Positive & 50\%  \\
& & False Positive  & 0\%  & & & False Positive  & 50\%    \\ \cline{2-4}\cline{6-8}

 & \multirow{2}{*} {\texttt{CallSuperInConstructor}} & True Positive & \cellcolor[HTML]{0350F8} 100\%& & \multirow{2}{*} {\texttt{CloseResource}}  & True Positive & 83.33\%  \\
& & False Positive  & 0\%  & & & False Positive  & 16.66    \\ \cline{2-4}\cline{6-8}

 & \multirow{2}{*} {\texttt{ClassNamingConventions}} & True Positive & 66.66\%& & \multirow{2}{*} {\texttt{EmptyWhileStmt}}  & True Positive & 75\%  \\
& & False Positive  & 33.33\%& & & False Positive  & 25\%      \\ \cline{2-4}\cline{6-8}

 & \multirow{2}{*} {\texttt{CommentDefaultAccessModifier}} & True Positive & \cellcolor[HTML]{0350F8} 100\% & & \multirow{2}{*} {\texttt{LoggerIsNotStaticFinal}}  & True Positive & 50\%  \\
& & False Positive  &0\%   & & & False Positive  & 50\%   \\ \cline{2-4}\cline{6-8}

 & \multirow{2}{*} {\texttt{ControlStatementBraces}} & True Positive & \cellcolor[HTML]{0350F8} 100\%  & & \multirow{2}{*} {\texttt{SuspiciousEqualsMethodName}}  & True Positive & 50\% \\
& & False Positive  & 0\%  & & & False Positive  & 50\%    \\ \cline{2-4}\cline{6-8}

 & \multirow{2}{*} {\texttt{DefaultPackage}} & True Positive & 50\%  & & \multirow{2}{*} {\texttt{UnnecessaryCaseChange}}  & True Positive & 75\% \\
& & False Positive  &50\%  & & & False Positive  & 25\%   \\\cline{2-4}\cline{6-8}
 & \multirow{2}{*} {\texttt{DontImportJavaLang}} & True Positive & \cellcolor[HTML]{0350F8} 100\%& & \multirow{2}{*} {\texttt{EmptyCatchBlock}}  & True Positive & 71.42\%  \\
& & False Positive  & 0\% & & & False Positive  & 28.57\%     \\ \cline{2-4}\cline{6-8}

 & \multirow{2}{*} {\texttt{EmptyMethodInAbstractClassShouldBeAbstract}} & True Positive & 60\%& & \multirow{2}{*} {\texttt{AvoidCatchingThrowable}}  & True Positive & \cellcolor[HTML]{0350F8} 100\%   \\
& & False Positive  & 40\%  & & & False Positive  & 0\%    \\ \cline{2-4}\cline{6-8}

 & \multirow{2}{*} {\texttt{FieldNamingConvention}} & True Positive & 0\%  & & \multirow{2}{*} {\texttt{AvoidInstanceofChecksInCatchClause}}  & True Positive & \cellcolor[HTML]{0350F8} 100\%  \\
& & False Positive  &\cellcolor[HTML]{FE0000} 100\%  & & & False Positive  & 0\%     \\ \cline{2-4}\cline{6-8}

 & \multirow{2}{*} {\texttt{FinalParameterInAbstractMethod}} & True Positive & 50\% & & \multirow{2}{*} {\texttt{CloneThrowsCloneNotSupportedException}}  & True Positive & \cellcolor[HTML]{0350F8}100\%   \\
& & False Positive  & 50\% & & & False Positive  & 0\%      \\ \cline{2-4}\cline{6-8}

 & \multirow{2}{*} {\texttt{ForLoopsMustUseBraces}} & True Positive & 87.5\% & & \multirow{2}{*} {\texttt{EqualsNull}}  & True Positive & \cellcolor[HTML]{0350F8} 100\% \\
& & False Positive  & 12.5\%  & & & False Positive  & 0\%     \\ \cline{2-4}\cline{6-8}

 & \multirow{2}{*} {\texttt{GenericsNaming}} & True Positive &\cellcolor[HTML]{0350F8} 100\% & & \multirow{2}{*} {\texttt{EmptyFinallyBlock}}  & True Positive & \cellcolor[HTML]{0350F8} 100\%  \\
& & False Positive  &0\%  & & & False Positive  & 0\%     \\ \cline{2-4}\cline{6-8}

 & \multirow{2}{*} {\texttt{IfElseStmtsMustUseBraces}} & True Positive & 90\%& & \multirow{2}{*} {\texttt{EmptyIfStmt}}  & True Positive & \cellcolor[HTML]{0350F8} 100\%  \\
& & False Positive  & 10\%  & & & False Positive  & 0\%   \\ \cline{2-4}\cline{6-8}

 & \multirow{2}{*} {\texttt{IfStmtsMustUseBraces}} & True Positive &\cellcolor[HTML]{0350F8} 100\%& & \multirow{2}{*} {\texttt{MissingBreakInSwitch}}  & True Positive & \cellcolor[HTML]{0350F8}100\%  \\
& & False Positive  &0\%  & & & False Positive  & 0\%   \\ \cline{2-4}\cline{6-8}

 & \multirow{2}{*} {\texttt{LocalVariableCouldBeFinal}} & True Positive & 66.66\% & & \multirow{2}{*} {\texttt{MissingStaticMethodInNonInstantiatableClass}}  & True Positive &  \cellcolor[HTML]{0350F8} 100\% \\
& & False Positive  & 33.33\%   & & & False Positive  & 0\%   \\ \cline{2-4}\cline{6-8}

 & \multirow{2}{*} {\texttt{LocalVariableNamingConventions}} & True Positive & \cellcolor[HTML]{0350F8} 100\% & & \multirow{2}{*} {\texttt{NonStaticInitializer}}  & True Positive & \cellcolor[HTML]{0350F8} 100\%   \\
& & False Positive  & 0\%  & & & False Positive  & 0\%   \\ \cline{2-4}\cline{6-8}

 & \multirow{2}{*} {\texttt{LongVariable}} & True Positive & 33.33\% & & \multirow{2}{*} {\texttt{NullAssignment}}  & True Positive & \cellcolor[HTML]{0350F8} 100\%   \\
& & False Positive  & 66.66\%  & & & False Positive  & 0\%     \\ \cline{2-4}\cline{6-8}

\bottomrule
\end{tabular}

\end{threeparttable}
}
\end{table*}

\begin{table*}
\caption{PMD rules that were perceived by students as true positive and false positive, broken down by category, as well as the percentages of responses for which students accepted/rejected the issues. Cells with 100\% acceptance are highlighted in blue, while cells with 100\% rejection are highlighted in red (Cont'd).}
\label{Table:pmd-tp-fp-both-b}
\resizebox{\textwidth}{!}{
 \begin{threeparttable}
\begin{tabular}{l|llc|l|lllccccc}
\toprule
\multicolumn{1}{l}{\textbf{Category}}& \multicolumn{1}{l}{\textbf{Rule}} & \multicolumn{1}{l}{\textbf{Metric}} & \multicolumn{1}{l|}{\textbf{Ratio}} & \multicolumn{1}{l}{\textbf{Category}}& \multicolumn{1}{l}{\textbf{Rule}} & \multicolumn{1}{l}{\textbf{Metric}} & \multicolumn{1}{l}{\textbf{Ratio}} \\ \hline

\parbox[t]{2mm}{\multirow{35}{*}{\rotatebox[origin=c]{90}{\textbf{Code Style (Cont'd)}}}} 
 & \multirow{2}{*} {\texttt{MethodArgumentCouldBeFinal}} & True Positive & 81.81\%& & \multirow{2}{*} {\texttt{ProperLogger}}  & True Positive & \cellcolor[HTML]{0350F8} 100\%   \\
& & False Positive  &18.18\%  & & & False Positive  & 0\%    \\ \cline{2-4}\cline{6-8}
 & \multirow{2}{*} {\texttt{MethodNamingConventions}} & True Positive & 60\%& \parbox[t]{2mm}{\multirow{12}{*}{\rotatebox[origin=c]{90}{\textbf{Error Prone (Cont'd)}}}} & \multirow{2}{*} {\texttt{OverrideBothEqualsAndHashcode}}  & True Positive & \cellcolor[HTML]{0350F8} 100\% \\
&  & False Positive  & 40\% & & & False Positive  &0\%     \\ \cline{2-4}\cline{6-8}

 & \multirow{2}{*} {\texttt{OnlyOneReturn}} & True Positive & 60\%& & \multirow{2}{*} {\texttt{ReturnEmptyArrayRatherThanNull}}  & True Positive &\cellcolor[HTML]{0350F8} 100\%   \\
& & False Positive  & 40\%  & & & False Positive  &0\%    \\ \cline{2-4}\cline{6-8}

 & \multirow{2}{*} {\texttt{ShortClassName}} & True Positive & 0\% & & \multirow{2}{*} {\texttt{ReturnEmptyCollectionRatherThanNull}}  & True Positive & \cellcolor[HTML]{0350F8} 100\%  \\
& & False Positive  & \cellcolor[HTML]{FE0000} 100\%   & & & False Positive  & 0\%   \\ \cline{2-4}\cline{6-8}

 & \multirow{2}{*} {\texttt{ShortMethodName}} & True Positive & 0\%& & \multirow{2}{*} {\texttt{UnconditionalIfStatement}}  & True Positive & \cellcolor[HTML]{0350F8} 100\%  \\
& & False Positive  & \cellcolor[HTML]{FE0000} 100\%   & & & False Positive  &0\%    \\ \cline{2-4}\cline{6-8}

 & \multirow{2}{*} {\texttt{ShortVariable}} & True Positive &80.95\% & & \multirow{2}{*} {\texttt{UseEqualsToCompareStrings}}  & True Positive & \cellcolor[HTML]{0350F8} 100\%  \\
& & False Positive  &19.04\%  & & & False Positive  &0\%     \\ \cline{2-4}\cline{6-8}

 & \multirow{2}{*} {\texttt{SuspiciousConstantFieldName}} & True Positive &\cellcolor[HTML]{0350F8} 100\%  & & \multirow{2}{*} {\texttt{AvoidMultipleUnaryOperators}}  & True Positive & 50\%  \\
& & False Positive  & 0\%  & & & False Positive  &   50\%
\\\cline{2-4}\cline{6-8}

 & \multirow{2}{*} {\texttt{TooManyStaticImport}} & True Positive & \cellcolor[HTML]{0350F8} 100\% & & \multirow{2}{*} {\texttt{DataflowAnomalyAnalysis}}  & True Positive & 0\%  \\ 
& & False Positive  &0\%   & & & False Positive  & \cellcolor[HTML]{FE0000} 100\%  \\ \cline{2-5} \cline{6-8} 

 & \multirow{2}{*} {\texttt{UnnecessaryConstructor}} & True Positive &\cellcolor[HTML]{0350F8} 100\% & \parbox[t]{2mm}{\multirow{45}{*}{\rotatebox[origin=c]{90}{\textbf{Performance}}}} & \multirow{2}{*} {\texttt{AddEmptyString}}  & True Positive & 75\% \\
& & False Positive  &0\%    & & & False Positive  & 25\%  \\ \cline{2-4}\cline{6-8}

 & \multirow{2}{*} {\texttt{UnnecessaryFullyQualifiedName}} & True Positive & \cellcolor[HTML]{0350F8}100\%  & & \multirow{2}{*} {\texttt{AvoidArrayLoops}}  & True Positive & 75\% \\
& & False Positive  &0\%   & & & False Positive  & 25\%  \\ \cline{2-4}\cline{6-8}

 & \multirow{2}{*} {\texttt{UnnecessaryImport}} & True Positive &\cellcolor[HTML]{0350F8}100\% & & \multirow{2}{*} {\texttt{AvoidFileStream}}  & True Positive &  \cellcolor[HTML]{0350F8} 100\%   \\
& & False Positive  & 0\%  & & & False Positive  & 0\%   \\ \cline{2-4}\cline{6-8}

 & \multirow{2}{*} {\texttt{UnnecessaryLocalBeforeReturn}} & True Positive & \cellcolor[HTML]{0350F8}100\% & & \multirow{2}{*} {\texttt{AvoidInstantiatingObjectsInLoops}}  & True Positive & 83.33  \\
& & False Positive  & 0\%  & & & False Positive  & 16.66    \\ \cline{2-4}\cline{6-8}

 & \multirow{2}{*} {\texttt{UnnecessaryModifier}} & True Positive &80\% & & \multirow{2}{*} {\texttt{AvoidUsingShortType}}  & True Positive &  \cellcolor[HTML]{0350F8} 100\%  \\
& & False Positive  &20\%   & & & False Positive  & 0\%    \\ \cline{2-4}\cline{6-8}

 & \multirow{2}{*} {\texttt{UnnecessaryReturn}} & True Positive & \cellcolor[HTML]{0350F8}100\% & & \multirow{2}{*} {\texttt{BooleanInstantiation}}  & True Positive &  \cellcolor[HTML]{0350F8} 100\%   \\
& & False Positive  &0\%   & & & False Positive  & 0\%    \\ \cline{2-4}\cline{6-8}

 & \multirow{2}{*} {\texttt{UseDiamondOperator}} & True Positive & \cellcolor[HTML]{0350F8}100\% & & \multirow{2}{*} {\texttt{ConsecutiveAppendsShouldReuse}}  & True Positive &  \cellcolor[HTML]{0350F8} 100\%   \\
& & False Positive  &0\%    & & & False Positive  & 0\%   \\ \cline{2-4}\cline{6-8}

 & \multirow{2}{*} {\texttt{UselessParentheses}} & True Positive & 70\%& & \multirow{2}{*} {\texttt{InefficientEmptyStringCheck}}  & True Positive &  \cellcolor[HTML]{0350F8} 100\%   \\
& & False Positive  &30\%    & & & False Positive  & 0\%    \\ \cline{2-4}\cline{6-8}

 & \multirow{2}{*} {\texttt{UseShortArrayInitializer}} & True Positive &0\%  & & \multirow{2}{*} {\texttt{InefficientStringBuffering}}  & True Positive & \cellcolor[HTML]{0350F8} 100\%  \\
& & False Positive  & \cellcolor[HTML]{FE0000} 100\%  & & & False Positive  & 0\%     \\ \cline{2-4}\cline{6-8}

 & \multirow{2}{*} {\texttt{VariableNamingConventions}} & True Positive &75\% & & \multirow{2}{*} {\texttt{InsufficientStringBufferDeclaration}}  & True Positive &  \cellcolor[HTML]{0350F8} 100\%  \\
& & False Positive  &25\%    & & & False Positive  & 0\%   \\ \cline{1-4}\cline{6-8}

\parbox[t]{2mm}{\multirow{8}{*}{\rotatebox[origin=c]{90}{\begin{tabular}[c]{@{}l@{}}\textbf{Documentation}\end{tabular}}}}
 & \multirow{2}{*} {\texttt{CommentRequired}} & True Positive & 90\% & & \multirow{2}{*} {\texttt{IntegerInstantiation}}  & True Positive & \cellcolor[HTML]{0350F8} 100\%   \\
& & False Positive  & 10\%   & & & False Positive  & 0\%   \\ \cline{2-4}\cline{6-8}

 & \multirow{2}{*} {\texttt{CommentSize}} & True Positive & 27.77\% & & \multirow{2}{*} {\texttt{LongInstantiation}}  & True Positive &  \cellcolor[HTML]{0350F8} 100\%  \\
& & False Positive  & 72.22\%    & & & False Positive  &0\%    \\ \cline{2-4}\cline{6-8}

 & \multirow{2}{*} {\texttt{UncommentedEmptyConstructor}} & True Positive &  \cellcolor[HTML]{0350F8} 100\% & & \multirow{2}{*} {\texttt{OptimizableToArrayCall}}  & True Positive &  \cellcolor[HTML]{0350F8} 100\%  \\
& & False Positive  & 0\%    & & & False Positive  & 0\%   \\ \cline{2-4}\cline{6-8}

 & \multirow{2}{*} {\texttt{UncommentedEmptyMethodBody}} & True Positive &  \cellcolor[HTML]{0350F8} 100\% & & \multirow{2}{*} {\texttt{RedundantFieldInitializer}}  & True Positive &  \cellcolor[HTML]{0350F8} 100\%  \\
& & False Positive  & 0\%    & & & False Positive  & 0\%    \\ \cline{1-4}\cline{6-8}

\parbox[t]{2mm}{\multirow{15}{*}{\rotatebox[origin=c]{90}{\begin{tabular}[c]{@{}l@{}}\textbf{Multithreading}\end{tabular}}}}
 & \multirow{2}{*} {\texttt{AvoidUsingVolatile}} & True Positive & 50\%& & \multirow{2}{*} {\texttt{SimplifyStartsWith}}  & True Positive &  \cellcolor[HTML]{0350F8} 100\%  \\
& & False Positive  & 50\%  & & & False Positive  & 0\%   \\ \cline{2-4}
\cline{6-8}

 & \multirow{2}{*} {\texttt{}} &  & & & \multirow{2}{*} {\texttt{StringInstantiation}}  & True Positive &  \cellcolor[HTML]{0350F8} 100\%   \\
& &   &    & & & False Positive  & 0\%    \\ 
\cline{6-8}

 & \multirow{2}{*} {\texttt{}} &  & & & \multirow{2}{*} {\texttt{TooFewBranchesForASwitchStatement}}  & True Positive & \cellcolor[HTML]{0350F8}100\%   \\
& &   &    & & & False Positive  & 0\%   \\ 
\cline{6-8}

 & \multirow{2}{*} {\texttt{}} &  & & & \multirow{2}{*} {\texttt{UnnecessaryWrapperObjectCreation}}  & True Positive &  \cellcolor[HTML]{0350F8} 100\%   \\
& &  &    & & & False Positive  & 0\%   \\ 
\cline{6-8}

 & \multirow{2}{*} {\texttt{}} &  & & & \multirow{2}{*} {\texttt{UseArrayListInsteadOfVector}}  & True Positive & 75\%  \\
& &   &    & & & False Positive  & 25\%   \\ 
\cline{6-8}

 & \multirow{2}{*} {\texttt{}} &  & & & \multirow{2}{*} {\texttt{UseIndexofChar}}  & True Positive &  \cellcolor[HTML]{0350F8} 100\%   \\
& &   &    & & & False Positive  & 0\%   \\ 
\cline{6-8}

 & \multirow{2}{*} {\texttt{}} &  & & & \multirow{2}{*} {\texttt{UselessStringValueOf}}  & True Positive &  \cellcolor[HTML]{0350F8} 100\%   \\
& &   &    & & & False Positive  & 0\%    \\ 
\cline{6-8}

 & \multirow{2}{*} {\texttt{}} &  & & & \multirow{2}{*} {\texttt{UseStringBufferForStringAppends}}  & True Positive &  \cellcolor[HTML]{0350F8} 100\%  \\
& &   &    & & & False Positive  & 0\%  \\ 
\cline{6-8}

\bottomrule
\end{tabular}

\end{threeparttable}
}
\end{table*}

\subsection{\RQone}


In Tables \ref{Table:pmd-tp-fp-both-a} and \ref{Table:pmd-tp-fp-both-b}, we illustrate PMD rules that are perceived by students as True Positive (TP) and False Positive (FP).  It is worth noting the diversity of
these warnings/violations, \ie they spread from warnings regarding style, code practice, and documentation, to warnings dealing with design and performance. Thus, upon analyzing students' assignment solutions, we cluster the issues according to the PMD ruleset categories listed in the PMD official documentation \footnote{https://pmd.sourceforge.io}, namely, `Best Practices', `Code Style', `Design', `Documentation', `Error Prone', `Multithreading', `Performance', and `Security'. These categories were captured at different levels of granularity (\eg package, class, method, and attributes). 
In the rest of this subsection, we provide a more in-depth analysis of these categories and the associated PMD rulesets.

\textbf{Category \#1: Best Practices.} This category refers to the rules which enforce generally accepted best practices that are vital in an overall assessment of software quality. This category assists students in uncovering code that might violate main design and coding strategies, or indicate areas that might consider unnecessarily inefficient or difficult to maintain. Examples of the rules perceived by students as true positives include \texttt{LooseCoupling}, \texttt{SwitchStmtsShouldHaveDefault}, and \texttt{JUnitAssertionsShouldIncludeMessage}. 

\textbf{Category \#2: Code Style.}
This category refers to the rules which enforce a specific coding style. Most prominently, brace and naming rules, consist of good coding practices regarding code blocks and naming conventions. This can be illustrated briefly by rules \texttt{IfStmtsMustUseBraces} and \texttt{ClassNamingConventions}, which shows that it should be followed by braces even if it is followed only by a single instruction and class names should be in camel case naming conventions to improve naming quality in the  code and reflect the actual purpose of the parameters and variables.

\textbf{Category \#3: Design.} This category refers to the rules which help discover design issues. Students captured design rules that contain best practices concerning overall code structure.  By way of illustration, \texttt{AvoidDeeplyNestedIfStmts} rule  indicates avoiding deeply nested if statements and \texttt{GodClass} shows the violation of single responsibility principles that increases the complexity of the code  (\eg \texttt{CyclomaticComplexity}).

\textbf{Category \#4: Documentation.} This category refers to the rules which are related to code documentation. Documentation is the description, in natural language, of the code-level changes and such description is crucial as it reveals the developer's rationale behind their coding decisions. \texttt{CommentRequired} rule is a good illustration of this group and shows that students seem to pay
attention to the quality of the code comments.

\textbf{Category \#5: Error Prone.}
This category refers to the rules which detect constructs that are either broken, extremely confusing, or prone to run time errors. One of the key aspects to avoid these issues is readability. If the students refactor the code to be easily read and understood, there is less chance for misunderstandings and coding mistakes, and so students spend less time comprehending code. This is exemplified by error-prone-related rules such as \texttt{AvoidDuplicateLiterals} and \texttt{MissingStaticMethodInNonInstantiatableClass}.

\textbf{Category \#6: Multithreading.} 
This category refers to the rules that flag issues when dealing with multiple threads of execution. For example, PMD recommends  \texttt{AvoidUsingVolatile} to be avoided as the keyword `volatile' is used to fine-tune a Java application and requires the expertise of the Java Memory Model.

\textbf{Category \#7: Performance.}
This category refers to the rules that flag sub-optimal code.
Since performance plays a vital role, students are encouraged to follow best coding standard practices such as \texttt{AvoidArrayLoops} and \texttt{AvoidInstantiatingObjectsInLoops} rules to avoid degrading the performance of the code.  

For our study, \textcolor{black}{we analyzed a total of 690 rules of students' selected issues} violating 155 distinct PMD rules. We found that 89 (57.41\%) issues had 100\% acceptance rate, while 9 (5.80\%) issues had 100\% rejection rate. The remaining had a mix of acceptance and rejection (36.79\%). As can be seen from Tables \ref{Table:pmd-tp-fp-both-a} and \ref{Table:pmd-tp-fp-both-b}, the majority of the violated rules are accepted and perceived by students as true positives, a few of these violated rules are rejected and perceived by students as false positives, and the remaining ones are distributed among both groups. While our results are not intended to be generalized, as it requires further experiments with larger sample sizes, our experience shows the success of PMD in triggering students' critical thinking about the problems outlined in the issues. Since students are given a choice to either accept or reject to address the issue, they can take the easy route of rejecting the majority of recommendations, as it is easier than accepting them and performing the necessary fixes. Fortunately, PMD has successfully attracted them to explore the various types of problems and provided sufficient documentation for them not only to comprehend it but also to code the needed fix for the problem. Moreover, it achieves another goal of our maintenance class, as it trains students to comprehend and act on code that they do not own.  

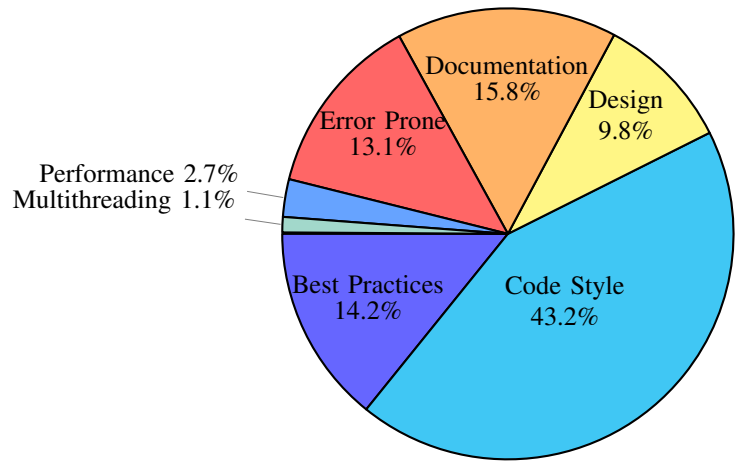
\begin{figure}[t]
\centering 
\begin{tikzpicture}
\begin{scope}[scale=1.0]
\pie[rotate = 180,pos ={0,0},text=inside,outside under=20,no number]{14.2/Best Practices\and14.2\%, 43.2/Code Style\and43.2\%, 9.8/Design\and9.8\%,15.8/Documentation\and15.8\%,13.1/Error Prone\and13.1\%, 2.7/Performance\and2.7\%, 1.1/Multithreading\and1.1\%}
\end{scope}
\end{tikzpicture}
\caption{Percentage of issues perceived as true positives and false positives, clustered by PMD categories.}
\label{fig:categories_clustered}
\end{figure}

Looking at the PMD rules that could play a role in acceptance and rejection, Figure \ref{fig:categories_clustered} depicts the percentages of issues perceived by students as both TP and FP. As can be seen, the most common PMD ruleset category perceived as TP and FP concerns `Code Style', representing 43.2\% of the issues. This observation is in line with the findings of previous studies describing that most code reviewers look for style conformance when evaluating the quality of code \cite{singh2017evaluating,gousios2015work}. The next most common categories are `Documentation', `Best Practices', and `Error Prone', representing 15.8\%, 14.2\%, and 13.1\% of the issues, respectively. This might indicate that students have different perspectives on whether developers follow the best practices, write descriptive enough code comments, or make code less susceptible to errors. The category `Design', `Performance', and `Multithreading' had the least number of issues perceived as TP and FP, which had a ratio of  9.9\%, 2.8\%, and 1.1\%, respectively.


\begin{tcolorbox}
\textit{Among the 690 analyzed issues, representing
155 distinct PMD rules. Nearly 75\% of the rules had 100\% acceptance rate, and only 6\% had 100\% rejection rate.} 
\end{tcolorbox}

\subsection{\RQtwo}

\begin{figure}[t]
 	\centering
 	\includegraphics[width=1.0\columnwidth]{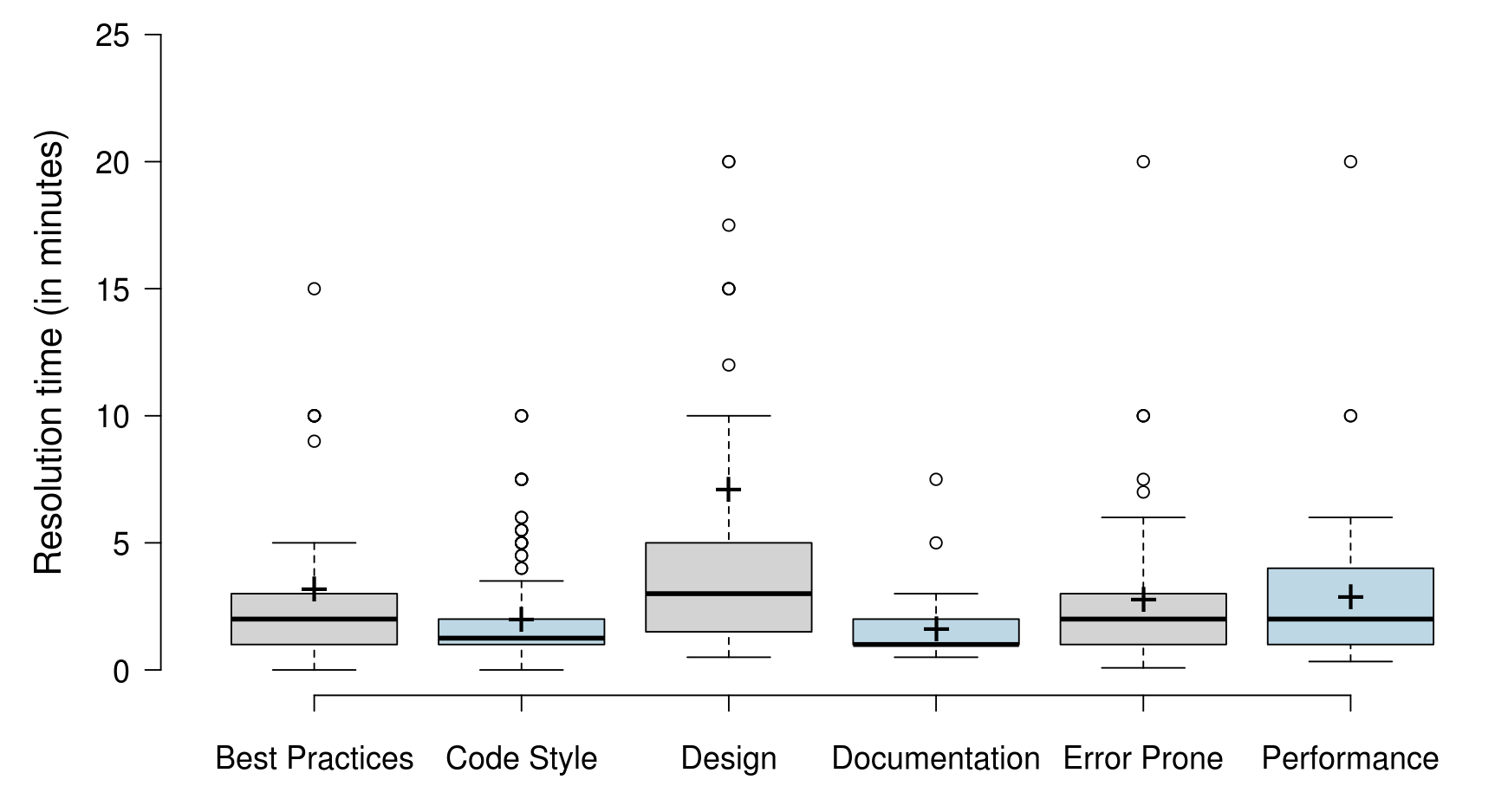}
 	\caption{Boxplots of time taken to fix issues, clustered by PMD ruleset categories.}
 	\label{fig:boxplot}
\end{figure}

By looking at Figure \ref{fig:boxplot}, we found issues belonging to the `Design' and `Best Practices' categories ($\mu$ = 7.11 and $\mu$ = 3.20, respectively) take more time to be fixed, compared to the other categories. We speculate that these two categories take time to fix as students need to apply techniques to resolve quality flaws or design issues, including code smells (\eg \texttt{GodClass}), and quality attributes (\eg \texttt{CyclomaticComplexity}). This is exemplified in the assignments undertaken by students to resolve the issues. Students refactor the code and optimize the design by performing repackaging, \ie extracting packages and moving the classes between these packages, and merging packages that have classes strongly related to each other. The present observations are significant in at least two major respects: (1) improving the quality of packages structure when optimizing cohesion, coupling, and complexity and (2) avoiding increasing the size of the large packages and/or merging packages into larger ones which might have an impact on the following design rules, namely, \texttt{CognitiveComplexity}, \texttt{CyclomaticComplexity}, and \texttt{NPathComplexity}. Further, students might find fixing PMD issues violating best coding practices to be a challenge  when trying to balance the trade-offs when maintaining efficient code with minimal errors. The other categories take less time to fix when comparing the mean (`Code Style' = 2.0, `Documentation' = 1.63, `Error Prone' = 2.79, `Performance' = 2.90), which might possibly hint at an easy resolution of the reported issues or warnings.

\textcolor{black}{
The fact that design issues take significantly longer to be resolved, reflects a deeper challenge for the students. These antipatterns are symptoms of poor design or architectural decisions, requiring students to comprehend the anticipated design first, in order to scope the symptoms of the antipattern. Unlike other issues, this requires going beyond one or few instructions, into reasoning over methods and classes, and how they are architected. While there is empirical evidence of how these antipatterns significantly increase the code’s proneness to errors \cite{khomh2009exploratory}, and hinder program comprehension \cite{bavota2015test}, there is no consensus on how to detect \cite{paiva2017evaluation} and correct them \cite{lacerda2020code}. Yet, students are not trained to handle the subjective nature of the problem, and therefore, it can potentially cause longer reflection before reaching a refactoring decision.}

\textcolor{black}{
Similarly to design patterns, being widely adopted in modeling classes \cite{designpatterns98,christensen2004frameworks,schulte2010introduction}, students should also be exposed to antipatterns. When design patterns are being taught, students are engaged in identifying key aspects of common design structure that make it reusable, while adhering to Object-Oriented design principles. Yet, existing large and complex systems are known to exhibit the existence of antipatterns \cite{palomba2018diffuseness}. Therefore, students shall be able to identify symptoms of bad design and programming practices, \ie \textbf{problem-based learning}. This learning paradigm leverages complex real-world problems (\eg antipatterns) to vehicle the learning of concepts (\eg design principles) \cite{duch2001power}. One success criterion of this paradigm is the ability of problems in motivating students to propose multiple solutions for their resolution. Antipatterns challenge students' design thinking as they reason over their correction. Since there are multiple refactoring opportunities associated with each antipattern, students would need to justify their choices. This can be achieved by measuring design quality, in terms of structural metrics, before and after the refactoring is implemented. Design evaluation is another aspect that students need to develop. Unlike error fixes, where students can systematically test their code for correctness, there is no trivial approach to validate the adequacy of a design change, without proactively measuring its impact on quality. The existence of multiple potential solutions, to address a given antipattern, can be leveraged by educators to establish a \textbf{cooperative learning} assignment. This paradigm allows students to compare their solutions through the assessment of refactorings' impact on antipattern resolution, along with its impact on software quality.}




\begin{tcolorbox}
\textit{Design' and `Best Practices' PMD ruleset categories take longer time to be resolved.}   

\end{tcolorbox}

\subsection{\RQthree}
\begin{table*}[ht]
  \centering
	 \caption{\textcolor{black}{Student's insight about the usefulness,  usability and functionality of the tool.}}
	 \label{Table:example}
\begin{adjustbox}{width=1.0\textwidth,center}
\begin{tabular}{llLllll}\hline
\toprule
\bfseries Category & \bfseries Sub-category & \bfseries Example (Excerpts from a related student's comment) \\
\midrule
\multirow{7}{*} {\textbf{Usefulness}} & Automation & \say{\textit{The PMD code plugin allows us to automatically run the PMD code analysis tool on our project’s source code and generate a site report with its results.}}\\
& \cellcolor{gray!30}   {Awareness} &  \cellcolor{gray!30} \say{\textit{It generates several defects with Id, and description
which will be helpful for developers to detect any bugs in source code.}} \\ 
& {Debugging} &  \say{\textit{The software is really powerful since it can find and categorize errors throughout the project very comprehensively and
systematically in categories, making the debugging process efficient and effective.}} \\ 
& \cellcolor{gray!30}  {Efficiency}  &  \cellcolor{gray!30}  \say{\textit{Bug detection and source code analysis are completed in a matter of seconds.}} \\ 
&  Quality & \say{\textit{it is great tool for keeping the code clean and maintainable.}} \\ 
\hline
\multirow{9}{*} {\textbf{Usability}} & Documentation & \say{\textit{PMD plugin provides a very detailed explanation about each and every bug
listed. It describes how it can be fixed along with an explanation why it was classified as an
issue.}} \\

&   \cellcolor{gray!30} Ease of use &  \cellcolor{gray!30} \say{\textit{The PMD is easy to use and covers a wider range code defects as compared to eclipse IDE. I 
have found it easy to use and the issue type are easy to understand.}} \\ 
&  Configurability & \say{\textit{I appreciated the custom
violations overview window - it clearly gives the PMD plugin moer flexibility over what information is
shown (and how the hierarchy is organized).}} \\ 
&   \cellcolor{gray!30} Visualization &   \cellcolor{gray!30} \say{\textit{I liked the fact that it had its own view within Eclipse to make it really easy what I’m looking at
and to organize all the different issues that were found. The labels and colors also made it
simple to determine the severity of anything that was found by the program so I knew what I
could look at and handle for this assignment reasonably.}} \\ 
\hline
\multirow{6}{*} {\textbf{Functionality}} & Design  &  \say{\textit{PMD provide a better introduction to 
the common programming flaws. It also provides possibilities for recognize the complex 
problems which insist many flaws.}} \\ 
& \cellcolor{gray!30} Rules & \cellcolor{gray!30} \say{\textit{I also appreciated the violations having very
specific, detailed names presented in a clearly standardized fashion. I do like the violation categories
(eg., Urgent, Important, Critical, Blockers, and Warnings), but the color scheme
and order of priority significance is not immediately apparent in the current design.}} \\
& Warning severities  & \say{\textit{A good feature is that they have filter for severity filter for these violations which help to focus on the critical ones.}} \\
\hline
\multirow{10}{*} {\textbf{Recommendation}} &  Automation &  \say{\textit{It is very simple and convenient to use. But there are still some shortcomings need manual analysis, but in general very good.}}  \\
& \cellcolor{gray!30} Correctness & \cellcolor{gray!30} \say{\textit{the reported bugs have two problems: First, the bugs sometimes repeat in reports,[...] maybe they are different at some detail in rules, but the performed result are identical. Another Problem is too much bugs about conventions in reported bugs. It’s quite excellent to make standardized name for variables, but in actual developments, especially informatization project, the name of variables should conform to the meaning, not to the conventions. Both of these two problems could be fixed with rules option, but for real use PMD needs more improvement.}}\\
&  Customization &  \say{\textit{PMD provides a clear list of issues per file, although there is no
straightforward method to group the issues by categories across the entire project.
It is relatively sound and works well for what it is.}}\\
& \cellcolor{gray!30} Resolution & \cellcolor{gray!30} \say{\textit{Need to provide more fixes as to how developers can proceed in fixing the bug. Found that to be lacking using this tool.}} \\

\bottomrule
\end{tabular}
\end{adjustbox}
\vspace{-.3cm}
\end{table*}

In Table~\ref{Table:example}, we report the main thoughts, comments, and suggestions about the overall impression of the usefulness, usability, functionality, and recommendation of the tool, in accordance with the conducted labeling. The table also presents samples of the students' comments to illustrate their impressions of each theme.


\textbf{Usefulness.} Generally, the respondents found the tool to be useful in regard to five main aspects: automation, awareness, debugging, efficiency, and quality. A group of students commented that the tool's ability to automatically perform code analysis on  a project’s source code and generate a report containing information about bugs is useful. Nearly 75\% of the students commented that PMD is intuitive to use and was efficient to locate software defects, and convenient for developers to use.  Further,  6.42\% of students commented that the tool's ability to find and categorize errors throughout the project makes the debugging process efficient and effective. A few students (12.84\%) revealed that the PMD was fast in terms of analyzing huge and complex software, and identifying \texttt{GodClass}, and \texttt{DataClass} was one of the perks. 5.5\% of students communicated that detecting the issues assists in increasing its \textit{readability}, which helps improve overall code quality. 

\textbf{Usability.}  Based on the feedback provided by the students, the key areas in usability related to documentation, ease of use, configurability, and visualization. 70\% of the students pointed out the tool is user-friendly and provides meaningful documentation with examples. Other comments also stated the tool's compatibility with various Java build tools, such as Maven, Ant, and Gradle, etc.

\textbf{Functionality.} According to the students' feedback about the tool's functional features, 60\% of the students' comments \textcolor{black}{appreciate the idea of the approach and are satisfied with various aspects of the tool’s operation}, and how this feature helps in better understanding of bad programming practices in real-world scenarios. Additionally, the students commented on their ability to practice a variety of strategies to remove issues. 20.18\% of students mentioned that they liked the rule violation categories, indicated by the tool, and the detailed descriptions of these violations as they helped them decide whether or not
the code should be fixed or if the detected issues were acceptable. 

\textbf{Recommendation.} From the students' feedback, we have also extracted suggestions to improve the tool. 28.3\% of the students' comments show a couple of suggested changes as a recommendation to be made to the tool's operation. We found out the students pointed out some of the recommendations related to automation, correctness, customization, and resolution. Students recommend the static analysis tool to be integrated with refactoring IDE instead of manually making those changes, as this feature will provide will allows developers to leverage built-in, safe refactoring features, instead of performing them manually. Other students felt that some issues are duplicates (\eg a student reported that \texttt{LiteralsFirstInComparisons} and \texttt{PositionLiteralsFirstInComparisons} are almost the same bug issues), and can be clustered, to reduce the number of suggested fixes. Students also recommended grouping the issues by categories across the entire project when outputting the analysis results, and generating informative messages and fixes (\eg  enable access to refactoring tools to resolve static analysis
warnings) since some of the flaws were observed with inexplicable fixes or suggestions. This observation is in line with previous studies \cite{christakis2016developers,johnson2013don} finding that bad warning messages and no suggested fixes are one of the pain points reported by industrial software developers when using program analyzers.

\begin{tcolorbox}
\textit{Overall, the students were satisfied with the PMD and rated its various aspects positively. 
} 
\end{tcolorbox}

\section{Reflection}
\label{Section:Reflection}
This section provides the lessons learned from both students' and educators' perspectives. Below, we first discuss our thoughts as to what went well, and what our plans are for Spring 2023. Then, we share the main reactions of the students.

\subsection{Student Perspective}

\noindent\textbf{Lesson \#1:  \textit{Learning the best coding practices for software quality improvement.}} Running a static analysis tool as an assignment allows students to learn best coding practices to improve code quality and make the code less vulnerable to errors. For example, upon analyzing students' comments about the tool, students learn and use refactoring strategies for code smell resolution and quality improvement.  Therefore, adopting the best practices and sharpening students' coding skills further improve development skills as a professional programmers, and support the characteristics of good code base health (\eg  maintainability, readability, and understandability).

\noindent\textbf{Lesson \#2:  \textit{Using the static analysis tool as a quick reference during the code review process.}} It is fundamentally vital to expose students to work on open-source projects to provide them with training to use the code that they did not write since upon their graduation, they are most likely to work on existing projects. That is why, this assignment challenges and trains them to perform code review, which is becoming a standard practice in the industry \cite{bacchelli2013expectations,sadowski2018modern,raibulet2019teaching,distefano2019scaling,sadowski2018lessons}. For instance, Google \cite{sadowski2018lessons} and Facebook \cite{distefano2019scaling} found that providing developers with an extensive list of warnings rarely motivates them to fix them, but reporting warnings during code review improves the adoption and removal of static analysis warnings. More specifically, students practice the review of different kinds of code changes (\ie functional and non-functional) as it helps a student to grow as a quality-aware developer.  \textcolor{black}{From a pedagogical perspective, code review is considered an active learning technique in SE education as it is a team-based activity and requires technical knowledge to review and analyze code \cite{hilburn2011read,chong2021assessing}. Our first and third research questions show how students are being challenged in code comprehending and fixing the issues. To cope with this challenge and develop students' critical and analytical skills, educators can consider applying \textbf{active learning} approach by following activities: (1) teach best practices for quality improvement, using metrics, and performing refactorings, (2) provide students with complementary tools for issues detection and correction, (3) instruct how to provide constructive criticism to others during code review, and (4) spread instructions of how to leave useful descriptive comments.} Furthermore, it is necessary to teach the next generation of software engineering students the best practices for reviewing code that can result in higher quality code since, so far, these skills are generally acquired through experience or training.


\subsection{Educator Perspective}


\noindent\textbf{Lesson \#1:  \textit{Creating custom PMD rules to enforce software engineering principles and good development practices.}} Using PMD was beneficial to students as it offers insights into various optimization possibilities and possible flaws in the code. According to the student's comments about the tool, students are interested in defining their own specific rules that would benefit their organization or future long-term class project. Thus, in the next course iteration, we plan to add scenarios where students are requested to design their own ruleset and use them to identify what they consider to be bad coding practices. Also, teachers can support students with crowd-sourced further PMD rules by mining repositories and detecting a range of faults in code provided on question-answering sites like Stack Overflow \cite{licorish2022combining}. Further, since research in code smells mentioned that existing approaches can be subjectively perceived by developers \cite{bryton2010reducing,mantyla2006subjective,di2018detecting,arcelli2016comparing,fontana2012automatic}, it is essential to translate that to students early enough so they learn how to customize static analysis tools, and know how to make their decision about their correction measure.

\noindent\textbf{Lesson \#2:  \textit{Developing complementary assignments.}} Finding that reviewing design-related code changes takes longer than other changes reaffirms the necessity of integrating existing tools and techniques that can assist students in the code review process. For this to be successful and not troublesome to the students, the static analysis assignments can be augmented with refactoring recommendations (\eg \textsc{JDeodorant} \footnote{\url{https://github.com/tsantalis/JDeodorant}}) and software metrics (\eg \textsc{Understand} \footnote{\url{https://scitools.com/}}) to help students with creating a pipeline of detecting issues, correcting them, and measuring the impact of their change in code quality. 
\textcolor{black}{Since one of our primary goals is to enhance students' problem-solving abilities, we rely on ASAT as a medium for \textbf{interactive learning}. When students analyze code, ASAT provides potential coding issues that violate coding standards. Therefore, students are being exposed to violations through examples, which facilitate their understanding. As students attempt their fix, they will interactively run the tool to verify the impact of their changes and close the feedback loop.} 
Moreover, we noticed that poorly naming the code elements is one of the main bad naming conventions practices, typically caught by students when reviewing code changes. Integrating chapters about naming convention, in the class, would support students with refactoring bad names. 


\noindent\textbf{Lesson \#3:  \textit{Training students for real-world setting.}} Students are typically given assignments where specific guidelines are given about how their work should address the outlined problems. Our assignment spins off by providing students with an open-ended problem, where they are given the freedom to select issues, and the responsibility to properly address them. It trains students to approach existing systems, and carefully choose their changes. Also, students learn how to justify their choices, either when accepting or denying \textcolor{black}{a given issue.}   \textcolor{black}{Furthermore, we observe from Figure \ref{fig:boxplot} that `Design' and `Best Practices' take longer to be resolved due to the fact that resolving quality issues might require effective correction measures. To facilitate the resolution time, increase students' engagement, and improve the team’s code review culture, we recommend educators implement \textbf{cooperative learning} strategy in the classroom in which students work in small groups to assist one another in learning the content. This can be achieved by applying the following tasks: (1) advocate for students to contribute to an open source project to fix issues as it has been shown that this helps with improving students' technical skills
and self-confidence \cite{pinto2019training}, (2) experience students with coming to a consensus during code review in cases opinions differ, and (3) engage students in early computing courses in the peer code review process.}



\section{Threats to Validity}
\label{Section:Threats}

In this section, we describe potential threats to the validity of our research method, and the actions we took to mitigate them.

\textbf{External Validity.} Concerning the generalizability of our
results, our study is limited to 65 submissions. Although we obtained valuable information and performed accurate analysis, the results may not represent the larger population of students that use static analysis tools. However, our participant pool is of a similar size (56) to the study that analyzed how industrial and open-source developers engaged with static analysis tools \cite{vassallo2020developers}. Further, our analysis was performed on mature open-source Java projects that varied in size, contributors, and number of commits. However, we cannot claim the generality of our observations to projects written in other programming languages or belonging to different ecosystems. Further investigation of even more projects is needed to mitigate this threat.

\textbf{Internal and Construct Validity.} As for the completeness and correctness of our interpretation of the open-ended comments about the tool, we did not extensively discuss all comments because some of them are open to various interpretations, and we need further follow-up interviews to clarify them. Additionally, to avoid personal bias during the manual analysis, each step in the manual analysis was conducted by two authors until reaching a consensus. The choice of PMD, as a static analysis tool, may introduce some bias to the way these issues are detected, especially since the detection of bad programming practices and code smells is known to be subjective \cite{bryton2010reducing,mantyla2006subjective,di2018detecting,arcelli2016comparing,fontana2012automatic}. Also, students may have had a different experience, if another tool was selected in this assignment. We chose PMD as it is one of the popular state-of-the-art tools, but in future work, we plan on trying other static analysis tools, to see if they can also reach this level of satisfaction.

Since students are choosing what to fix, they may skip fixing relevant warnings for other non-technical reasons (\eg late assignment submission). However, since the rejection ratio is low, we believe that students did their best to take the issues seriously.

\section{Conclusion}
\label{Section:Conclusion}
Understanding the practice of reviewing code to improve the quality is
of paramount importance to education. Although modern code review is widely adopted in open-source and industrial projects, \textcolor{black}{the relationship between the usage of ASATs such as PMD and how students perceive it during code analysis remains unexplored. In this study, we performed a quantitative and qualitative study to explore the effectiveness of PMD in familiarizing students with improving source code, by i) detecting code issues and antipatterns, and ii) implementing fixes for their correction. The paper develops the culture of reviewing  and patching unknown code.} 

Our results reveal that several kinds of ASAT warnings that students pay more attention to during code review, reviewing design and best practices related changes take longer to be completed compared to other changes, and students rated various aspects of the tool positively, while also providing valuable ideas for future development. For future work, we plan on using other ASATs which will complement and validate our current study to provide the software engineering community with a more comprehensive view of the use of ASATs in order to engage students with software quality improvement from educator and student perspectives. Moreover, we plan to investigate students' understanding of code review practice using various real-world applications in a semester-long course project.

\bibliographystyle{ieeetr}
{\scriptsize\bibliography{references}}
\end{document}